\documentstyle[12pt]{article}

\newlength{\dinwidth}
\newlength{\dinmargin}
\def\lapproxeq{\lower .7ex\hbox{$\;\stackrel{\textstyle
<}{\sim}\;$}}
\def\gapproxeq{\lower .7ex\hbox{$\;\stackrel{\textstyle
>}{\sim}\;$}}

\begin{document}
\titlepage
\begin{flushright}
RAL-TR-95-013 \\
DTP/95/48  \\
June 1995 \\
\end{flushright}

\begin{center}
\vspace*{2cm}
{\Large {\bf The $\alpha_S$ Dependence of Parton Distributions}}
\\

\vspace*{1cm}
{\large A.\ D.\ Martin and W.\ J.\ Stirling,} \\
\vspace*{0.25cm}
{\it Department of Physics, University of Durham, \\
Durham, DH1 3LE, England} \\

\vspace*{0.5cm}
and \\

\vspace*{0.5cm}
{\large R.\ G.\ Roberts,} \\
\vspace*{0.25cm}
{\it Rutherford Appleton Laboratory, Chilton, Didcot, \\
Oxon, OX11 0QX, England.}
\end{center}

\vspace*{3cm}
\begin{abstract}
We perform next-to-leading order global analyses of deep
inelastic and related data for different fixed values of
$\alpha_S (M_Z^2)$.  We present sets of parton distributions for
six values of $\alpha_S$ in the range 0.105 to 0.130.  We display
the $(x, Q^2)$ domains with the largest parton uncertainty and we
discuss how forthcoming data may be able to improve the
determination of the parton densities.
\end{abstract}

\newpage

Analysis of the scaling violations of deep inelastic scattering
data provides one of the accurate ways to determine the QCD
coupling $\alpha_S$.  The scaling violations observed in recent
high precision muon and neutrino deep inelastic data yield
values\footnote{Throughout we work in terms of the QCD
coupling at the $Z$ pole, $Q^2 = M_Z^2$, evaluated in the
$\overline{\rm MS}$ scheme with 5 flavours, which we shall denote
simply by $\alpha_S$.} of $\alpha_S = 0.113 \pm 0.005$ \cite{mv}
and $0.111 \pm 0.006$ \cite{cc} respectively.  Moreover a global
parton analysis which includes these data, with other related
data, gives $\alpha_S = 0.113 \pm 0.005$ \cite{mrsa}.  However,
there are other independent determinations which lie outside this
range; for example $\alpha_S$ determined from LEP event shapes or
from $\tau$ decays.  A recent review of all the methods is given
by Webber \cite{brw} who concludes that the world average is
$0.117 \pm 0.005$.

$\alpha_S$ is now also being determined from jet rates observed
in the experiments at HERA \cite{h1a} and Fermilab \cite{fnal}.
These methods have the advantage of determining $\alpha_S$ over a
wide range of $Q^2$ and it is believed that they will eventually
yield values with equal precision to the other determinations of
$\alpha_S$.  However,
they require the use of parton distributions which have their own
particular value of $\alpha_S$ and so the question of consistency
arises.  Does the \lq\lq output" $\alpha_S$ depend on the \lq\lq
input" value of $\alpha_S$?  In order that the sensitivity to the
input partons may be studied, we repeat the global analysis of
refs.\ \cite{mrsa,mrsg} for various fixed values of $\alpha_S
(M_Z^2)$ in the interval which covers these other independent
determinations, namely 0.105 to 0.130.  This study allows us to
highlight the deep inelastic data that particularly constrain
$\alpha_S$.  Moreover it provides a quantitative estimate of the
uncertainty associated with the parton distributions $f_i (x,
Q^2)$ in different regions of $x$ and $Q^2$.

An earlier analysis \cite{mrse}, which studied the uncertainty in
the determination of $\alpha_S$, did provide 4 parton sets which
cover a limited range of $\alpha_S$.  Here we extend the range
and use the improved set of deep inelastic and related data.
CTEQ \cite{cteq} have recently presented an additional parton set
with a high $\alpha_S$, and Vogt \cite{v} has provided 5 sets of
GRV \cite{grv} partons with $\alpha_S$ in the interval (0.104,
0.122).  The latter is not a global analysis and so cannot
accommodate the variation of partons, particularly at larger $x$,
which attempt to compensate for the shift of $\alpha_S$ from its
optimum value.

We base our global next-to-leading analysis on the MRS(A)
parametric forms, updated in ref.\ \cite{mrsg} to include recent
HERA data \cite{h93,z93}.  That is we consider variations about
the optimum fit MRS(A$^\prime$)\footnote{We choose not to base
our analysis on the MRS(G) set of partons \cite{mrsg}, since this
would involve an extra parameter which is only relevant to data
at very small $x$.  It therefore would introduce a degree of
ambiguity in a domain where the precision of the data is going to
rapidly improve.  We return to this point later.}.  Fig.\ 1 shows
the $\chi^2$ values for various subsets of deep inelastic data
obtained in six new global fits\footnote{The six sets have
$\alpha_S = 0.105, 0.110, \ldots 0.130$ and are denoted by
MRS.105 etc.  The parton sets can be obtained by electronic mail
from W.J.Stirling@durham.ac.uk.} with different values of
$\alpha_S (M_Z^2)$.  Cross section and asymmetry data for
Drell-Yan and $W$ hadroproduction are included in the analysis,
but their contributions to $\chi^2$ are not shown.  These data
remain well-described in the fits with different $\alpha_S$
values by slight adjustments of the partonic flavour structure of
the proton; they do not pin down $\alpha_S$.  Also the data for
prompt photon production are accommodated as $\alpha_S$ varies by
a change in scale.  The $\chi^2$ profile for the WA70 data is
shown in Fig.\ 1.  Only the smallest values of $\alpha_S$ are
disfavoured; the prediction does not fall off quite steeply
enough, as the photon transverse momentum increases, to agree
with the observed distribution.

The HERA 1993 data \cite{h93,z93} are included in the analysis.
In addition we show the $\chi^2$ values for the preliminary 1994
ZEUS data \cite{z94} that were obtained with the electron-proton
collision point shifted (so that the detectors can gather deep
inelastic events at smaller $x$).  Due to the logarithmic scale
that has been used for $\chi^2$ it is easy to be misled by Fig.\
1 about the relative importance of various data sets in the
determination of $\alpha_S$.  The $\chi^2$ profiles at the top of
the plot have a more significant impact than those which lie
lower down.

Considerable insight into the effect of varying $\alpha_S$ (and
the related ambiguities) can be obtained from Fig.\ 2.  This
shows the available data for $F_2^{ep} = F_2^{\mu p}$ at three
particular $x$ values:  $x = 0.0008$ in the HERA range, $x =
0.05$ which is relevant for $W$ production at Fermilab and $x =
0.35$ representative of the large $x$ BCDMS precision data which
provide the tightest constraints on $\alpha_S$.  The curves are
obtained from the three parton sets which have $\alpha_S = 0.105,
0.115$ and 0.125.  Recall that the optimum overall description
occurs for $\alpha_S = 0.113$ and so the continuous curve gives a
better global fit than the ones either side.  As expected, the
scaling violation is greatest for the partons with the largest
value of $\alpha_S$.  Also, as may perhaps be anticipated, the
curves cross in the region of the data, which lie in different
intervals of $Q^2$ for the different values of $x$.  Away from
the $(x, Q^2)$ domain of the data the predictions show a
considerable spread.  For example for $x = 0.0008$ and $Q^2 \sim
10^3 \: {\rm GeV}^2$ we see quite a variation in the prediction
for $F_2^{ep}$.  The ambiguity in the small $x$ domain is
actually greater than that shown, since the quark sea and the
gluon have been constrained to have the same small $x$ behaviour,
that is
\begin{equation}
x S \sim A_S x^{- \lambda_{\cal S}}, \: \: \: xg \sim A_g x^{-
\lambda_g}
\label{a1}
\end{equation}
\noindent with $\lambda_S = \lambda_g$.  Unfortunately HERA is
unable to measure $F_2$ in this region of $x$ and $Q^2$; the
reach of the collider is kinematically limited to the domain
$x/Q^2 \gapproxeq 10^{- 5} \: {\rm GeV}^{-2}$.  Nevertheless as
the precision of the HERA data improves it will be possible to
determine $\lambda_S$ and $\lambda_g$ independently (see
\cite{mrsg}).  The sensitivity of the predictions to the
interplay between the form of the gluon and the value of
$\alpha_S$ demonstrates the importance of a global analysis which
includes the crucial large $x$ constraints on $\alpha_S$.  The
$\chi^2$ profiles shown in Fig.\ 1 that are obtained from the
HERA data overconstrain $\alpha_S$ since they are based on fits
which set $\lambda_S = \lambda_g$.  In our global \lq\lq
$\alpha_S$" analysis this has a negligible effect on the partons,
except at small $x$, where for $Q^2$ values away from
the HERA data there will be more variation than the spread that
our curves imply.

The lower plot in Fig.\ 2 shows a typical set of the high
precision BCDMS data \cite{bcdms}.  In the large $x$ domain these
data place tight constraints on $\alpha_S$, free from the
ambiguity associated with the gluon.

Fig.\ 3 gives another view of the constraints on the partons in
various regions of $x$ and $Q^2$.  It is a contour plot of the
ratio $R$ of $F_2$ as predicted by two sets of partons with
$\alpha_S (M_Z^2)$ fixed either side of the optimum value.  In
$(x, Q^2)$ regions of precise data, acceptable fits demand that
the ratio $R$ be equal to 1.  This is strikingly borne out.  We
see that the $R = 1$ contour lies precisely in the centre of the
band of the fixed-target deep inelastic scattering data.  Again
$R \simeq 1$ in the region of the more accurate HERA data, that
is $x \lapproxeq 10^{- 3}$ and $Q^2 \sim 15 \: {\rm GeV}^2$.
When the precision of the HERA data improves and the accuracy
extends over a larger domain of $x$ and $Q^2$ we would expect the
$R \simeq 1$ contour to also track the HERA band, but probably at
the expense of allowing $\lambda_S$ and $\lambda_g$ to be free
independent parameters.  Of course Fig.\ 3 is just an overview of
the description of one observable $(F_2^{ep} = F_2^{\mu p})$.
The global fit has many other constraints to satisfy.
Nevertheless $F_2^{ep}$ is measured over far wider regions of $x$
and $Q^2$ than the other observables and so Fig.\ 3 gives a
useful indication of features of the global fit.

Additional experimental data in regions where the contours in
Fig.\ 3 are closely spaced would clearly have significant impact
on the analysis.  For example sufficiently precise information in
the region of high $x$ and low $Q^2$ (the lower left corner of
Fig.\ 3) would help pin down $\alpha_S$ even further.  This is
the domain of the SLAC experiments \cite{slac}.  However we must
take care to avoid regions where there are appreciable
target-mass/higher-twist effects.  For the SLAC data these
effects are smallest in the region $x \sim 0.3$, provided that
$Q^2 \gapproxeq 5 \: {\rm GeV}^2$ \cite{mv,mv2,mrsq}.  The Table
below shows the $\chi^2$ values for the subset of the SLAC data
\cite{slac} that lie in the \lq\lq safe" region ($0.18 \lapproxeq
x \lapproxeq 0.45$) obtained from our sets of partons with
different $\alpha_S$:

\begin{center}
\begin{tabular}{|c|c|c|c|c|c|c|}
\hline
$\alpha_S(M_Z^2)$ & 0.105 & 0.110 & 0.115 & 0.120 & 0.125 & 0.130
\\
\hline
$\chi^2$(25 pts) & 37 & 24 & 25 & 46 & 93 & 179 \\
\hline
\end{tabular}
\end{center}

\noindent The SLAC data clearly support the optimum value of
$\alpha_S$ determined by the other deep inelastic data, that is
$\alpha_S$ = 0.113.  The inclusion of the SLAC data in the global
fits would simply mean that the curves at $x=0.35$, for example,
would cross-over at a value just below the $Q^2 = 50 \: {\rm
GeV}^2$ intersection shown in Fig.\ 2.

Returning to Fig.\ 3 we see that jet production at large $E_T$ at
Fermilab samples partons in an $x,Q^2$ domain far removed from
the regions where deep inelastic measurements exist.  For
example, jets produced centrally with a transverse energy $E_T =
300$ GeV sample $x=2E_T/\sqrt{s} \sim 0.3$ and $Q^2 \sim E_T^2$.
Fig.\ 4 compares the observed single-jet inclusive distribution
of CDF \cite{cdf} with the predictions from partons with three
different values of $\alpha_S$.  For simplicity we have evaluated
the leading-order expression at a common renormalization and
factorization scale $\mu = E_T/2$.  Of course a precision
comparison between data and theory will require a full
next-to-leading order analysis \cite{ho}.  However, our aim here
is to compare the spread of the predictions with the uncertainty
of the data.  A leading-order calculation is entirely sufficient
for
this purpose.  A change of scale simply boosts the predictions up
or down relative to the data, but leaves the shapes in Fig.\ 4
essentially unchanged.  To consider the implications for partons
it is necessary to discuss the description in two distinct
regions of $E_T$.  For $E_T \lapproxeq 200$ GeV the jet
cross section is dominated by the $gg$ and $qg$ initiated
subprocesses.  Here the cross section ratios reflect the
different shapes of the gluon distributions in the region
$0.05 \lapproxeq x \lapproxeq 0.2$, with the predictions spread
out even more by the differences in the associated values of
$\alpha_S^2$.  For $E_T \gapproxeq 200$ GeV, on the other hand,
the cross section becomes increasingly dominated by
quark-initiated subprocesses.  Here the differences between the
curves reflect the spread in the predictions of $F_2$ at large
$x$ ($x \gapproxeq 0.25$) and large $Q^2$ ($Q^2 \sim E_T^2$), but
now suppressed (rather than enhanced) by the differences in
$\alpha_S^2(Q^2)$.  This illustrative exercise demonstrates the
value of a precise measurement of the jet distribution.  If the
experimental uncertainties can be reduced then these data will
impose valuable constraints on the gluons at small $x$ ($x \sim
0.1$) and on the quarks at large $x$ ($x \sim 0.35$), as well as
on the value of $\alpha_S$.

In summary we note that deep inelastic scattering data determine
$\alpha_S$ to be $0.113 \pm 0.005$.  This value is found in an
analysis of the BCDMS and SLAC data by Milsztajn and Virchaux
\cite{mv} and in the global analyses \cite{mrsa,mrsg} which
include, besides the BCDMS measurements, other deep inelastic and
related data.  Moreover the SLAC deep inelastic data \cite{slac},
which are not included in the global fit, also favour this value
of $\alpha_S$, see the Table above and the sample data in Fig.\
2.  The deep inelastic determination of $\alpha_S$ relies mainly
on the scaling violations of the data in the {\it large} $x$
domain $(x \gapproxeq 0.2)$, a region free from the gluon and its
ambiguities.  It is easy to verify that the low $x$ HERA deep
inelastic data for $F_2^{ep}$ do {\it not} determine $\alpha_S$
unless restrictive assumptions about the small $x$ behaviour of
the gluon and sea quark distributions are made.  Indeed we find
that the HERA data give little constraint on $\alpha_S$ even with
the assumption that $\lambda_S = \lambda_g$ in (\ref{a1}).  The
values of $\alpha_S$ determined from other (non deep-inelastic)
processes cover a wider interval:  $0.110 \lapproxeq \alpha_S
\lapproxeq 0.125$ \cite{brw}.  Some methods rely on input partons
and so there is a need for parton sets with values of $\alpha_S$
which cover this interval.  We have therefore performed a series
of global analyses of the deep inelastic data to obtain realistic
sets of partons corresponding to a sequence of values of
$\alpha_S (M_Z^2)$.  Since $\alpha_S$ is not optimal these are
compromise fits, which yield curves that intersect in the $x,
Q^2$ regions where precise data exist; see, for example, Fig.\ 2.
The body of the deep inelastic data occurs in the region $Q^2
\simeq 20$ GeV$^2$.  Fig.\ 5 shows the gluon and up quark
distributions from 3 parton sets with different $\alpha_S$ for
$Q^2$ values above and below this value.  The systematics
displayed in these plots may be anticipated from Fig.\ 2.  For
low $Q^2$, below the body of the data, we see from Fig.\ 5(a)
that the lower $\alpha_S$ partons \lq\lq swing more about the $x
\sim 0.05 - 0.1$ pivot points in favour of lower
$x$"; and vice-versa for the high $Q^2$ partons of Fig.\ 5(b).
It is interesting to note that $W$ and $Z$ boson production at
the Fermilab $p\overline{p}$ collider sample $u$ and $d$ partons
with $x \sim 0.05$ and $Q^2 \sim 10^4$ GeV$^2$, that is in the
region of the pivot point.  The predicted production cross
sections are therefore unusually stable to the change of the set
of partons used in the calculation\footnote{In particular we find
that there is only a spread of $\pm 3\%$ between the $W$ cross
section predictions obtained from our parton sets with $\alpha_S
= 0.115 \pm 0.010$, which is well below the present experimental
uncertainty, see, for example, ref.\ \cite{mrsa}.}.  Jet
production, on the other hand, samples partons over a range of
$x$ and $Q^2$, as well as being directly dependent on $\alpha_S
(Q^2)$.  It is therefore important to have to hand sets of
realistic partons with different $\alpha_S (M_Z^2)$.

\bigskip
\noindent {\large \bf Acknowledgements}

\medskip
We thank Nigel Glover for useful discussions and Joel Feltesse,
Mark Lancaster and Anwar Bhatti for information concerning the
data.

\newpage

\noindent {\large \bf Figure Captions}

\begin{itemize}
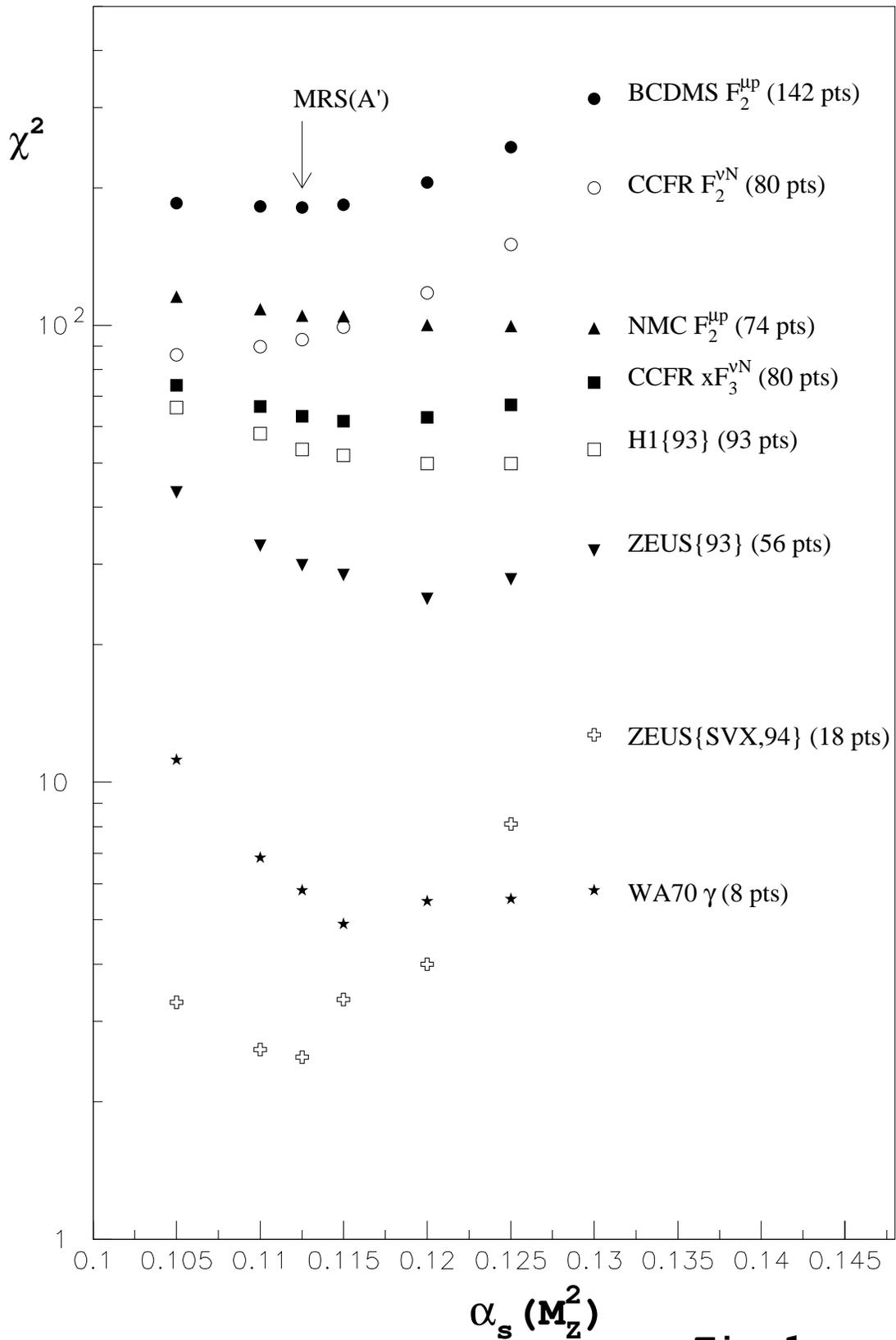
\item[Fig.\ 1] The $\chi^2$ values versus the value of $\alpha_S$
used in the global analysis.  The contributions to $\chi^2$ are
shown for the BCDMS \cite{bcdms}, CCFR \cite{ccfr}, NMC
\cite{nmc}, WA70 \cite{wa70}, H1 (1993) \cite{h93} and ZEUS
(1993) \cite{z93} data sets.  The $\chi^2$ values for the
preliminary ZEUS (SVX) 1994 \cite{z94} data are also shown, but
these data are {\it not} included in the fit.  The $\chi^2$
values are also shown for the MRS(A$^\prime$) set of partons
\cite{mrsg}, which has the optimum value of $\alpha_S$, namely
$\alpha_S = 0.113$.

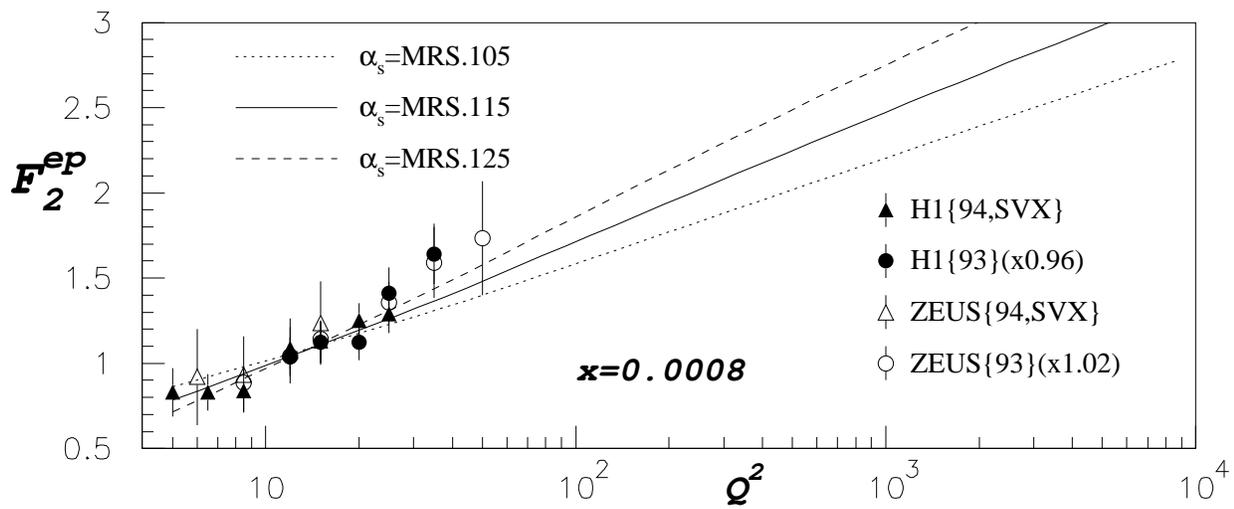
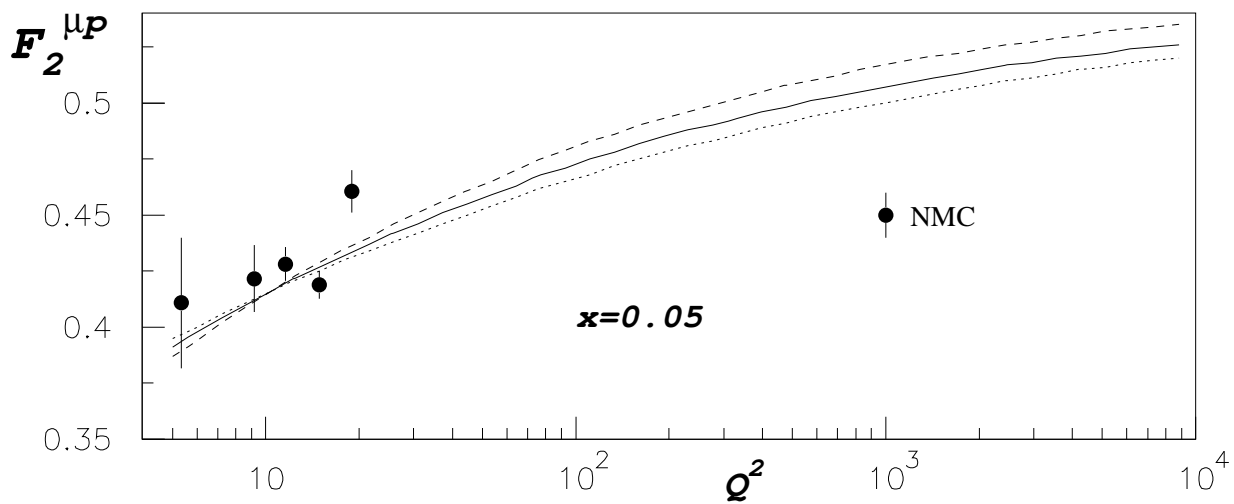
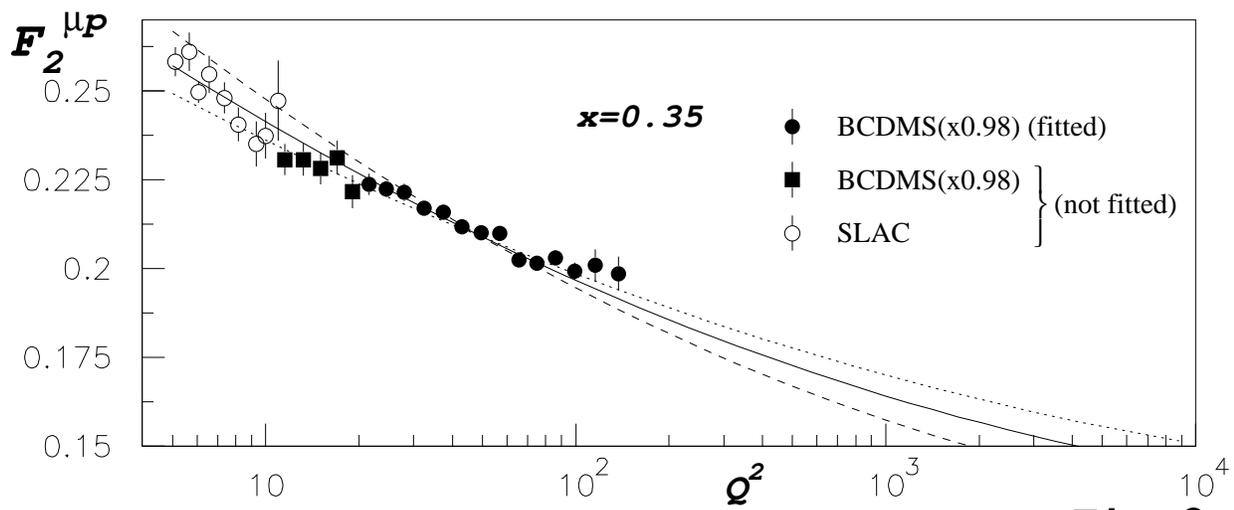
\item[Fig.\ 2] The scaling violations of $F_2^{ep} = F_2^{\mu p}$
at three different values of $x$.  The data are from refs.\
\cite{bcdms,nmc,h93,z93,z94,h194}.  The curves correspond to the
global parton fits with $\alpha_S = 0.105, 0.115$ and 0.125.
At $x=0.35$ some data points not normally included in our global
fits are also shown: first, the lower $Q^2$ BCDMS measurements
which are made only at their lower beam energy and, second,
SLAC data \cite{slac} in the region which is insensitive to
target-mass/higher-twist corrections.

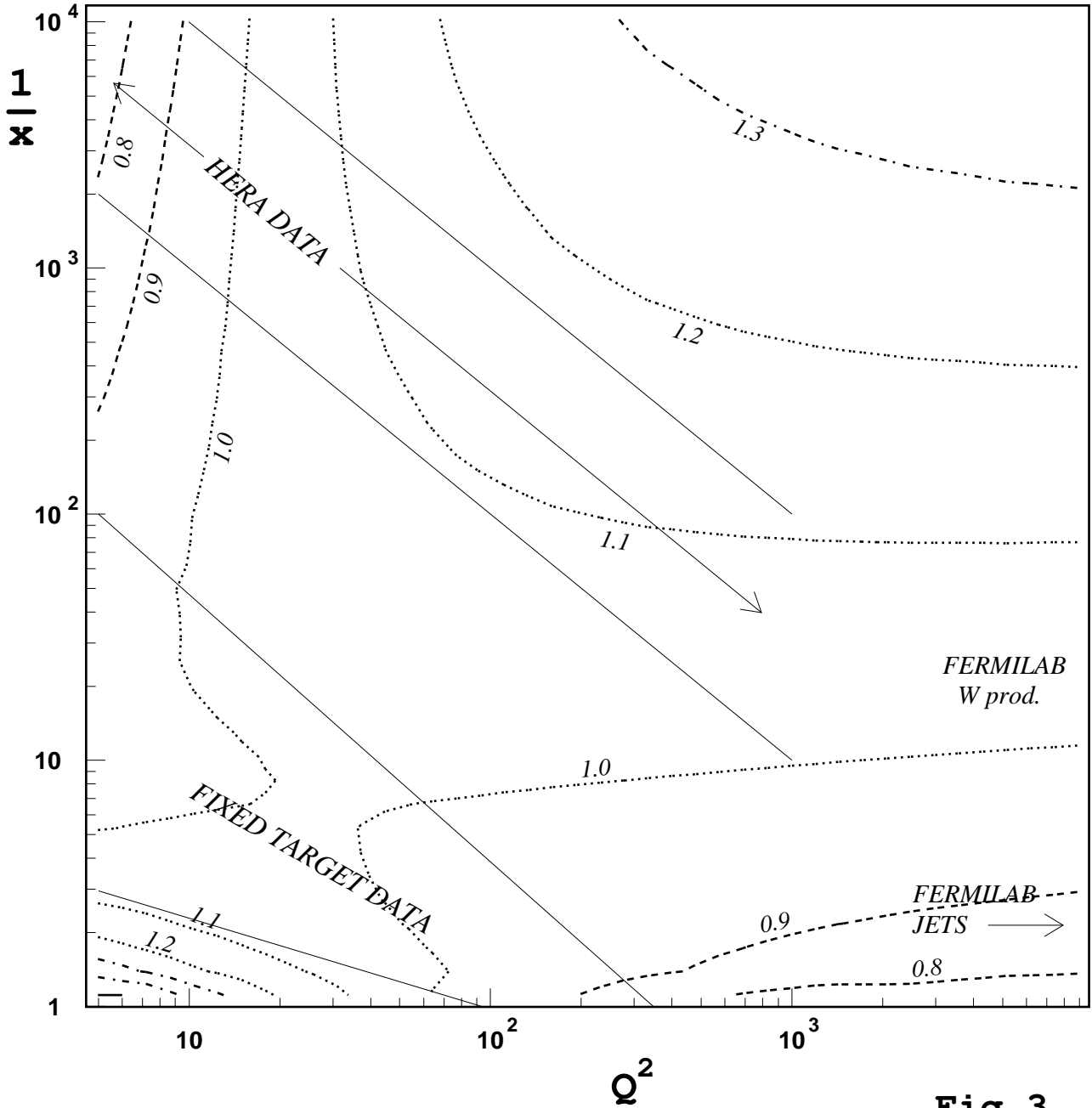
\item[Fig.\ 3] Contours of fixed $R \equiv F_2^{ep} (\alpha_S =
0.125)/F_2^{ep} (\alpha_S = 0.105)$ in the $x, Q^2$ plane, where
$F_2^{ep} (\alpha_S)$ is the structure function calculated from
partons obtained in a global analysis with $\alpha_S$ fixed at
the given value.  The bands indicate the regions where
measurements of $F_2$ exist.  The HERA data are much more precise
towards the low $Q^2$ end of the band.

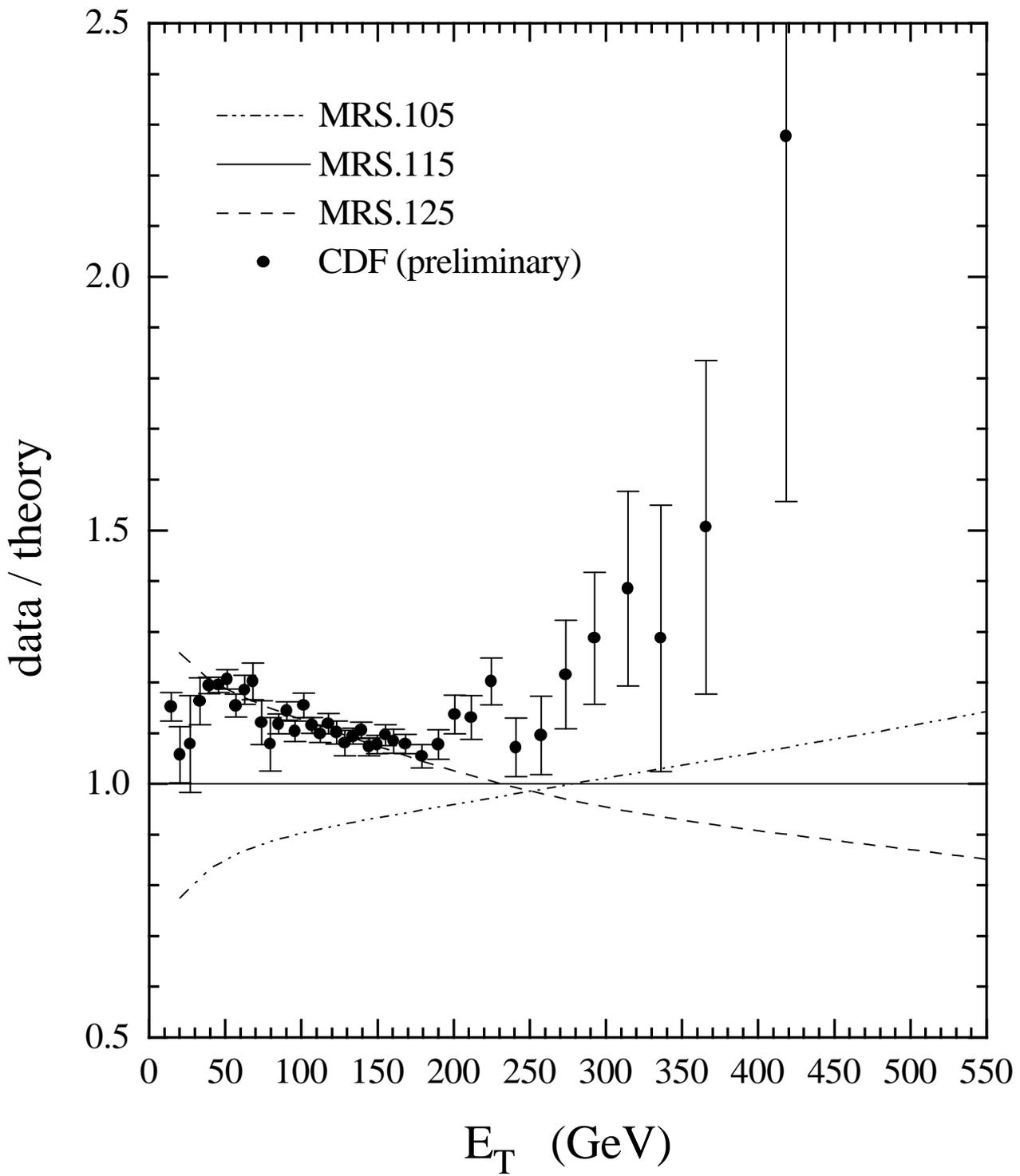
\item[Fig.\ 4] The $p\bar p$--initiated jet $E_T$ distribution at
$\sqrt s$ = 1.8 TeV normalized to the prediction from partons
with $\alpha_S = 0.115$ (i.e. MRS.115).  The data are the CDF
measurements of $d^2 \sigma/dE_T d \eta$ averaged over the
rapidity interval $0.1 < |\eta| < 0.7$ \cite{cdf}.  The curves
are
obtained from a leading-order calculation evaluated at $\eta =
0.4$.  The data are preliminary and only the statistical errors
are shown.  The systematic errors are approximately 25\% and are
correlated between different $E_T$ points.  We thank the CDF
collaboration for permission to show these data.

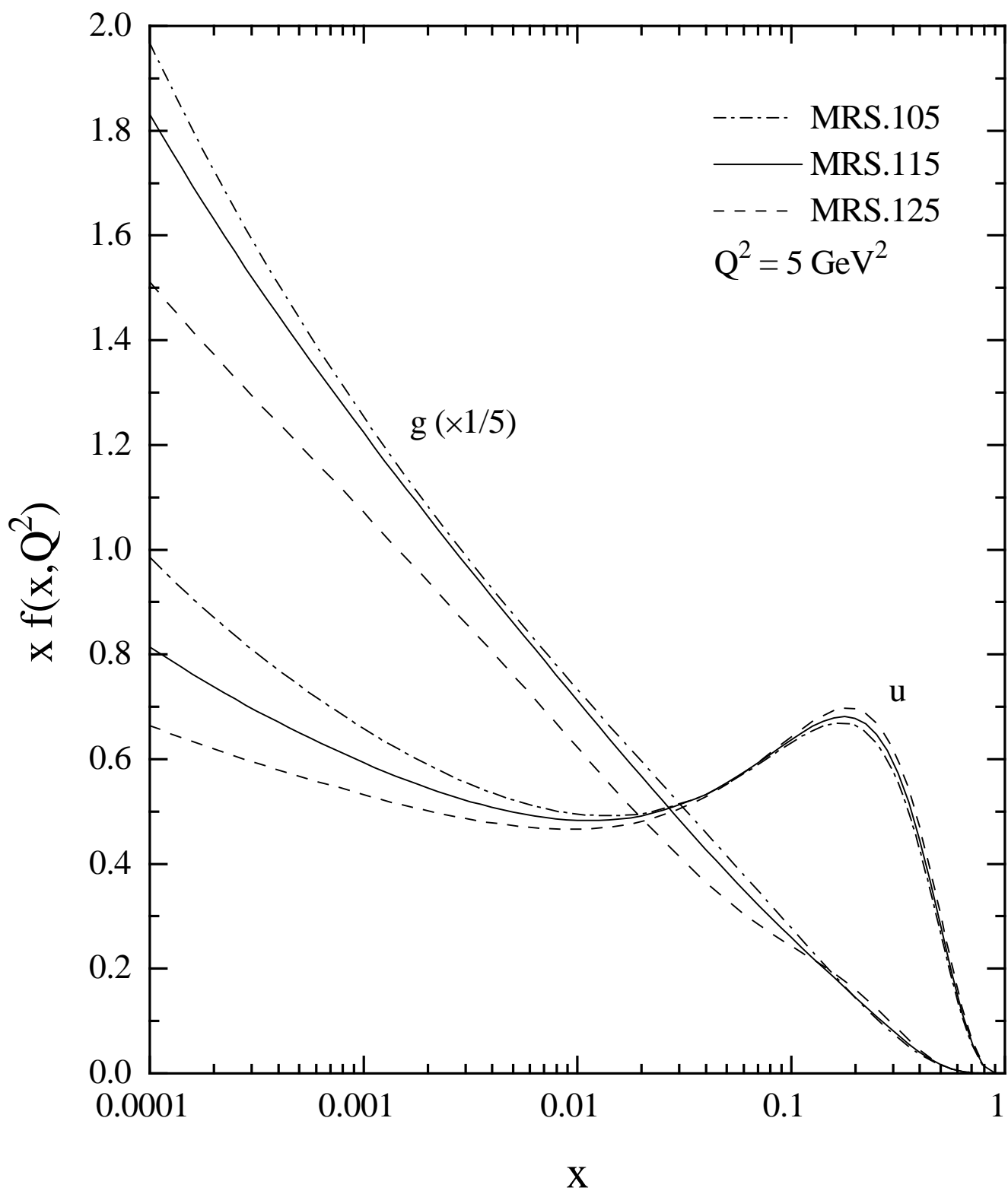
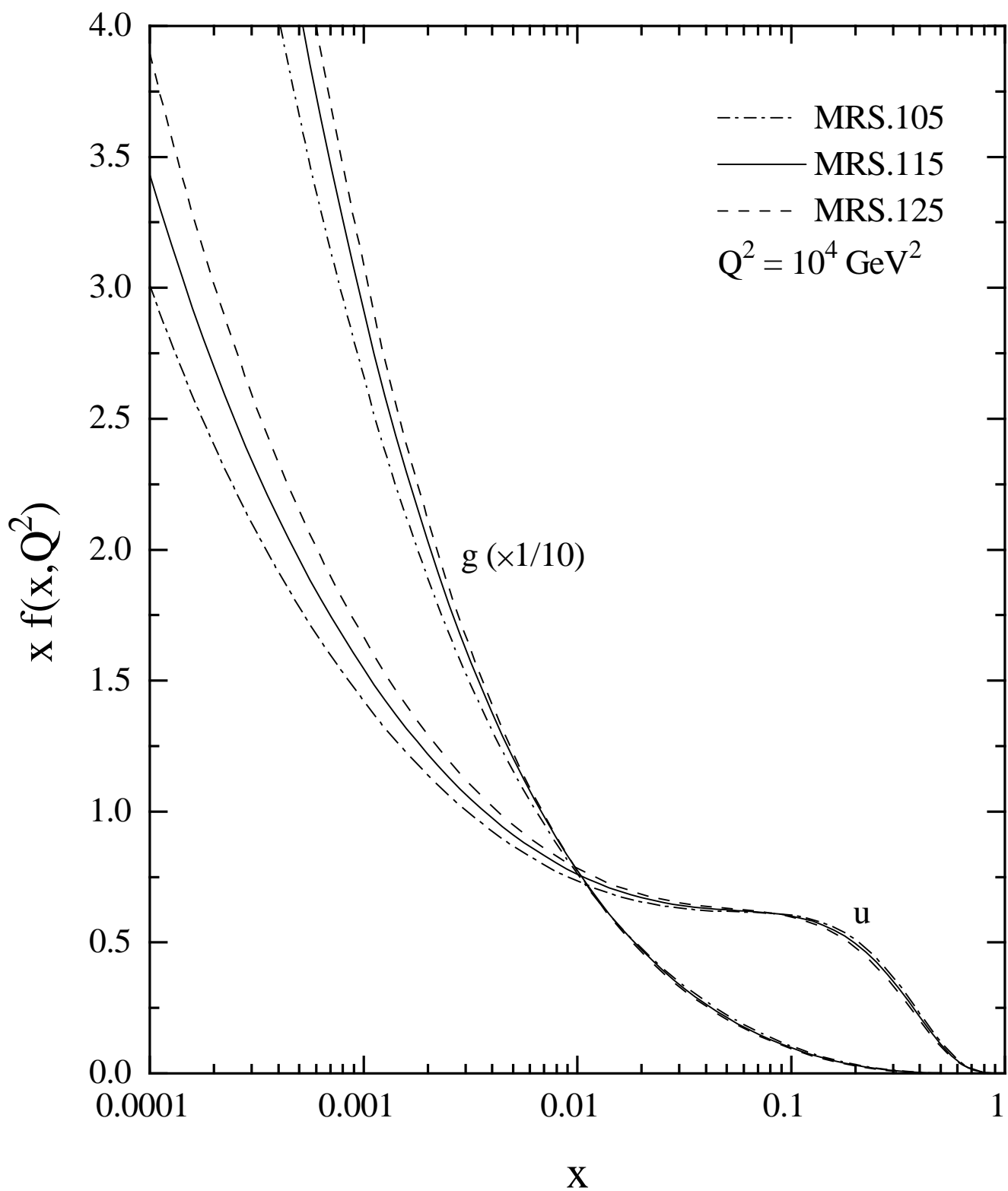
\item[Fig.\ 5] The $xg$ and $xu$ parton distributions at (a) $Q^2
= 5$ GeV$^2$ and (b) $Q^2 = 10^4$ GeV$^2$ of the parton sets with
$\alpha_S = 0.105$, 0.115 and 0.125.

\end{itemize}

\end{document}